# Reaction-Drift Model for Switching Transients in $Pr_{0.7}Ca_{0.3}MnO_3$-Based Resistive RAM

A. Khanna, S. Prasad, N. Panwar, and U. Ganguly

***Abstract*—Earlier, the DC hole-current modeling of PCMO RRAM by drift-diffusion (DD) including self-heating (SH) in TCAD (but without ionic transport) was able to explain the experimentally observed SCLC characteristics, prior to resistive switching. Further, transient analysis using DD+SH model was able to reproduce the experimentally observed fast current increase at ~100ns timescale followed by saturation increases, prior to resistive switching. However, resistive switching requires the inclusion of ionic transport. We propose a Reaction-Drift (RD) model of oxide ions, which is combined with the DD+SH model. Experimentally, SET operations consist of 3 stages and RESET operations consists of 4 stages. The DD+SH+RD model is able to reproduce the entire transient behavior over $10^{-8} - 1s$ range in timescale for both SET and RESET operations for a range of bias, temperature. Remarkably, a universal RESET behavior of $log(I) \propto m * log(t)$, where $m \approx -1/10$, is reproduced. The quantitatively different voltage time dilemma for SET and RESET is also replicated for a range of ambient temperature. This demonstrates a comprehensive model for resistance switching in PCMO based RRAM.**

## I. Introduction

Non-filamentary $Pr_{0.7}Ca_{0.3}MnO_3$ (PCMO) based resistive switching memory devices (RRAMs) are attractive due to better variability and multi-level resistance states [1], [2]. A forming-less operation is observed in PCMO [3], [4], which simplifies the memory controller. From a mechanisms perspective, in PCMO based RRAM, the following extent of understanding exists in literature. Qualitatively, Space Charge Limited Current (SCLC) mechanism has been invoked for current transport [5], [6], [7]. Resistance is modulated by trap-density – consistent with trap SCLC [8]. We have presented a simple trap density extraction methodology based on trap SCLC model to correlate trap density change with resistance switching [9]. Further, a TCAD model consisting of drift-diffusion (DD) based band of holes transport in p-type semiconductor with self-heating (SH) (but without ionic transport) to model SCLC current is able to replicate experimental dc IV characteristics at lower bias, i.e. prior to onset of resistive switching for a range of ambient temperatures ($25°C - 125°C$) [10]. The inclusion of self-heating enabled the replication of non-linear behavior, earlier erroneously attributed to Trap-Filled Limit [8]. The signature of self-heating was further confirmed by fast (sub-100ns) transient switching behavior [11]. Further, transient TCAD modeling was able to match the experimental current transient *prior* to the onset of resistance switching. Though, resistive switching behavior involving ion dynamics has not been modeled yet [10], [11], a *qualitative* explanation of resistive switching in PCMO based RRAM is as follows. Resistive switching in PCMO based RRAM is related to the transport of oxygen ion (or equivalently oxygen vacancies) [12], [13], [14], [15]. Reversible ionic transport occurs by reversing bias polarity to drift ions to and from a reactive electrode (i.e. an oxygen source/sink) to modulate oxygen vacancy concentration in PCMO [16], [17], [18]. These oxygen vacancies are related to hole traps [9]. Thus, ionic transport modulates trap concentration to produce resistance modulation of current under trap SCLC mechanism [8]. While such a qualitative model has been presented, the detailed dynamics of SET/RESET needs to be explored and quantitatively modeled. Recently, we have experimentally studied the long range ($10^{-8} - 1$ s) transient to highlights the signature of ion dynamics for SET and RESET as shown in Fig. 1. Three stages in SET operation were observed due to a step bias input - (S1) initial current increase is followed by (S2) current saturation, which is followed by (S3) abrupt current increase to compliance. Similarly, a four-stage RESET is observed where (R1) an initial increase in current is followed by (R2) a fast current decrease to (R3) a current saturation level. Finally, (R4) a slow universal current reduction of $I \propto t^{-1/10}$ is observed over 6 orders of magnitude in time. Such a specific and multi-stage behavior is attractive for quantitative model development to demonstrate detailed understanding of the switching mechanism.

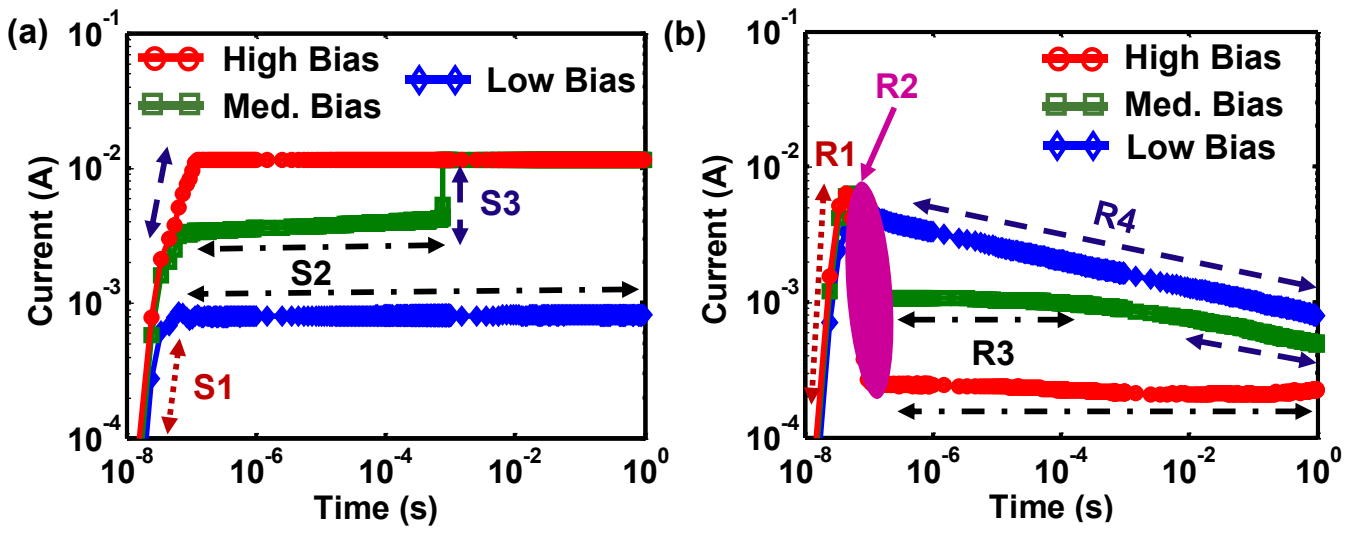

Fig. 1. Experimental (a) SET (b) RESET Transient is shown at various applied bias. For SET, a 3 stage transient characteristic is observed from a step voltage input - (S1) initial current increase due to self-heating is followed by (S2) current saturation, which is followed by (S3) abrupt current increase to compliance. The final current shoot-up to compliance is due to ion-transport. For RESET, the 4 stage transient characteristics is observed from a step voltage input - (R1) an initial current increase in current is followed by (R2) a fast current decrease to (R3) a current saturation level. Finally, (R4) a slow universal current reduction of $I \propto t^{-1/10}$ , over 6 orders of magnitude in time

In this paper, we introduce an ionic Reaction-Drift (R-D) model to include ion dynamics, coupled with our prior TCAD based DD+SH model [10]. We show that the model quantitatively replicates the experimental current transients for a range of bias including at different ambient temperature ($300 - 450K$). Thus, such an analysis is able to provide a quantitative understanding of resistive switching mechanism in PCMO based RRAM.

## II. Transient Resistive Switching & Model

Earlier, the DD+SH model (without ionic transport/reaction) was able to capture the short timescale transient response (stage S1), which was dependent upon self-heating timescale, to saturation (stage S2) until ionic transport occurs for SET/RESET where current increases sharply (stage S3). The deviation of simulations from experiment due to ionic transport is as follows. For SET, ionic motion leads to a current increase, which increases Joule heating in the device (i.e. self-heating) to further increase ionic motion. Thus, a positive feedback mechanism is set up to create a sudden sharp increase in current to compliance. As ionic transport was not included earlier, the essential resistance change dynamics was not modeled. The ionic transport, indicated by the ionic drift velocity ($v_{drift}$) at a given temperature and electric field, is given by Mott-Gurney Equation,

$$v_{drift} = a \times f \times exp\left(-\frac{E_m}{kT}\right) sinh\left(\frac{\xi}{\xi_0}\right) \qquad (1)$$

where a is hopping distance, f is escape frequency, ξ is an electric field, $\xi_o = kT/qa$ is characteristic electric field and $E_m$ is activation barrier, k is Boltzmann constant, a is hopping distance, T is absolute temperature. Ionic motion timescale is essentially the timescale at which $v_{drift}$ produces significant ionic motion. As the current is measured in log scale from 10ns to 1s, when the measurement timescale matches the ionic motion timescale for a given applied bias, the positive feedback is observed as a sudden sharp take off in current towards compliance (stage S3). As bias is reduced, we observe an exponentially longer timescale in saturation (stage S2) to abrupt current take-off (stage S3) i.e. occurrence of ion motion. This is qualitatively consistent with (1). Essentially, lower field and lower current cause lower heating and consequently lower temperature. Lower electric field (ξ) and temperature (T), will reduce $v_{drift}$ to increase timescale for equivalent ionic transport distance exponentially as shown in (1). To validate this qualitative explanation, a quantitative model based on numerical simulations is presented in this paper.

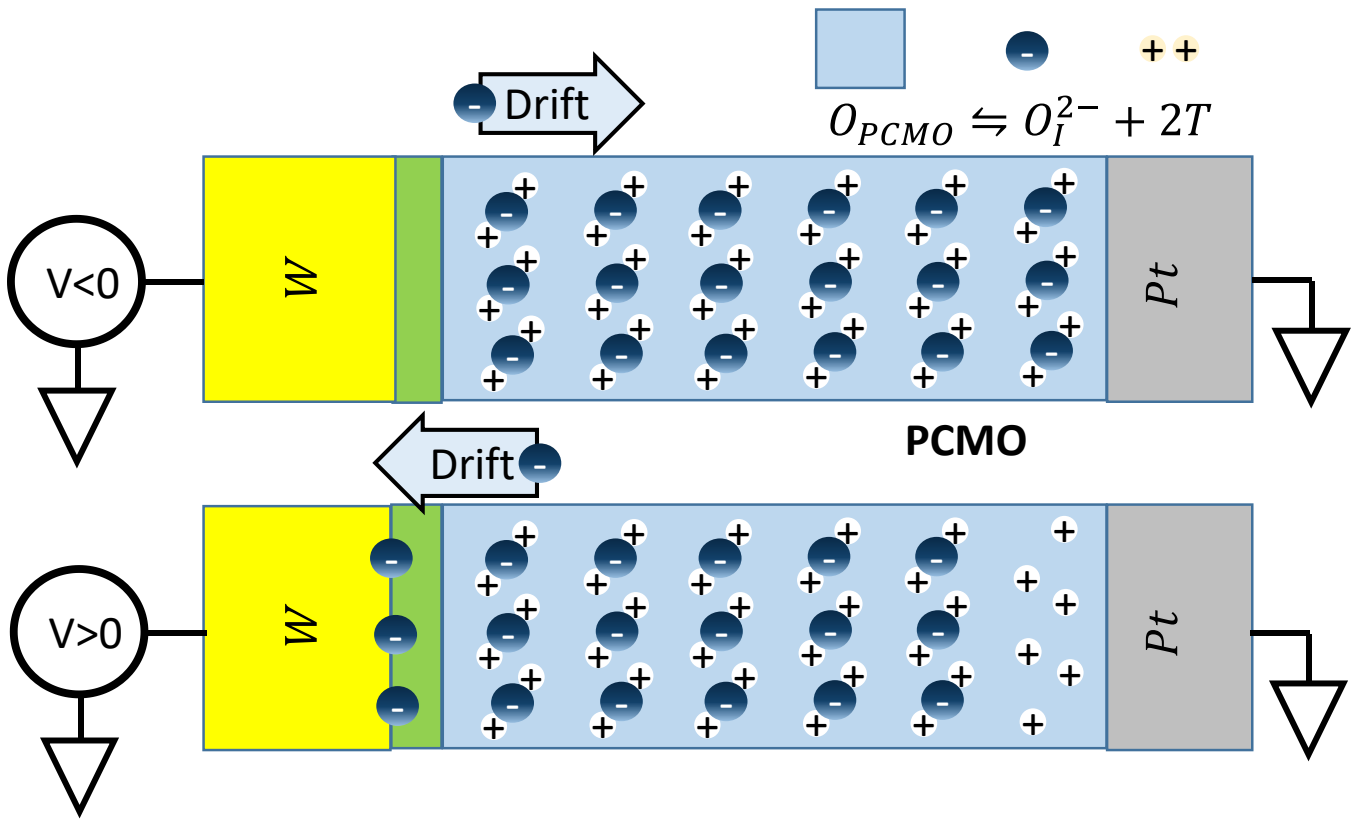

Fig. 2. Schematic view of generation and movements of ions and vacancies inside the PCMO device in presence of applied electric field. The oxide ions

The RESET transient phenomenon has 3 different applied voltage regimes. The high bias has a fast increase (stage R1) and then fast decrease in current (stage R2) followed by a long time-independent current (stage R3). For intermediate bias, the high bias behavior is observed, but the constant current (stage R3) is higher. Eventually, the constant current starts to reduce to follow the "universal" curve with the time exponent of approximately $-1/10$ (stage R4). The low bias shows initial current increase (stage R1) followed by the "universal" current transient curve with the time exponent of $-1/10$ (stage R4). We first model this *specific* "universal" behavior.

First, we show that a "universal" power law dependence is possible in an isothermal case as shown in Negative Bias Temperature Instability (NBTI) analysis in MOSFETs [19], [20] with insufficient (i.e. larger than observed) time exponent. For RESET, an increase in positive trap density reduces current [9], [10] as given in (2).

$$I_{trapSCLC} \sim \frac{I_{trapfree}}{\frac{N_T}{N_V}\exp(\frac{E_T - E_V}{kT})} \propto \frac{1}{N_T} \qquad (2)$$

We assume a three-step process. First, trap and interstitial anions are created, which is fast and hence close to equilibrium. To create positive traps, a substitutional unit element ($A_{lattice}$) produces an anion ($A^{n-}$) with n charges and a vacancy that yield n traps ($h_T{}^{+}$).

$$A_{lattice} \rightleftharpoons A^{n-} + n \quad ^{+} \qquad (3)$$

Second, bias polarity based drift is essential to explain bipolarity in bipolar RRAM. Under applied electric field, $[A^{n-}]$ can drift towards the reactive electrode. Third, when A reaches the reactive electrode, it is consumed e.g.

$$W + xA = WA_x \qquad (4)$$

We assume that this three-step process is drift limited, i.e. the reaction (2) essentially close to equilibrium.

$$n\frac{d[h_T]}{dt} = \frac{d[A^{n-}]}{dt} = k_F[A_{lattice}] - k_R[A^{n-}][h_T]^n$$

$$constant = k_F[A_{lattice}] \approx k_R[A^{n-}][h_T]^n \qquad (5)$$

$$[A^{n-}] \approx k[h_T]^{-n} \qquad (6)$$

Also, any $A^{n-}$ ions reaching the reactive electrode is instantaneously consumed. Total amount of $A^{n-}$ per unit area is given by $[A^{n-}] * L$, where L is the device length. Hence the rate of remove of $A^{n-}$ is dependent upon drift velocity ($v_{drift}$) as given below, where we assume $v_{drift}$ to be position independent for simplicity

$$n\frac{d[h_T] * L}{dt} = \frac{d[A^{n-}] * L}{dt} = v_{drift} * [A^{n-}] \qquad (7)$$

Using (6), we get,

$$n\frac{d[h_T] * L}{dt} = \frac{v_{drift}k}{[h_T]^n} \qquad (8)$$

Integrating, we get

$$\frac{nL}{n+1} * [h_T]^{n+1} = v_{drift}kt \qquad (9)$$

The time evolution of trapped hole concentration ($[h_T]$) is given by

$$[h_T] = \left(\frac{(n+1)v_{drift}k}{\text{nL}}\right)^{\frac{1}{n+1}} t^{\frac{1}{n+1}} \qquad (10)$$

Assuming $[h_T] = N_T$ i.e uniform trap density, then by (1), we get

$$I_{trapSCLC} \propto t^{-\frac{1}{n+1}} \qquad (11)$$

Thus, a single anion (A) creating n trap based reaction-drift (RD) model will produce a time exponent $m = 1/(n+1)$. The experimental exponent of $\approx -1/10$ requires $n = 9$. Oxygen is the only anion in PCMO. This is indeed quite difficult that a single diffusing species of oxygen cause 9 traps. The first possibility is an oxide ion ($O^{2-}$), which should produce $n = 2$ traps is widely reported which is produce an exponent of $m =$ 1/3. Next, a superoxide ion ($O_2^-$) should produce n = 4 traps. However, superoxide ions are reported for surface diffusion and dissociation into $O^{2-}$ ions for bulk diffusion in solid oxide fuel cell (SOFC) electrode studies of LSMO [21]. Thus, for an isothermal case, it is difficult to imagine a reaction where n = 9.

Next, we show that qualitatively self-heating will reduce the time exponent of the power law. For RESET, the device starts in low resistance state. Upon application of bias, the current increases with self-heating (stage R1). Consequently, temperature rises and ions transport are initiated to increase trap density, which reduces the current (stage R2). First, in the case of high bias, a very high current is reached to enable a high temperature quickly. At high temperature, fast ion transport occurs to reduce current before the stored heat escapes i.e. temperature can drop. Once temperature drops, ionic transport stops. Trap density and consequently current becomes time independent. Second, in the case of lower bias, a lower final current is reached (stage R1). Consequently, the temperature rises to a moderate level, where ionic transport increases trap density slowly to reduce the current (stage R4). The gentle current reduction will reduce the temperature gently, which will reduce the ionic transport. A negative feedback occurs due to temperature reduction, which will slow down ionic transport in time compared to the isothermal case. Naturally, the rate of current reduction will also slow down. Thus, the power law exponent will reduce from $m = -1/3$. We will show that we can achieve the exponent of $m \approx -1/10$ using detailed numerical simulations. Based on this model, we show also that high bias regime i.e. fast increase and then fast decrease of current followed by constant current is also realized. Further, the intermediate bias regime i.e. initially high bias like behavior (stage R1 & R2), which results in a higher constant current will remain unchanged in time (stage R3) as the initial log-timescale of measurement is too fast for ionic transport at that intrinsic device temperature due to self-heating. However, at longer timescale of measurement when the timescale becomes comparable to the rate of ionic transport at that intrinsic temperature, then the transient current reduces again and merges with the "universal" curve of time exponent $m = -1/10$ (stage R4).

## III. Simulations Set Up

In the next section, we develop the numerical simulation bench, which will simultaneously be able to reproduce experimental SET and RESET transient response at various ambient temperatures.

### A. Initialization

We start with the device structure as given in the experiment. The device is initially in LRS state defined by low uniform $N_T(t=0)$ and an initial temperature ($T(t=0)$), which is essentially ambient temperature. For RESET switching, low uniform $N_T(t=0)$ (initially in LRS state) is assumed whereas for SET high uniform $N_T(t=0)$ (initial in HRS state) is assumed – which may be extracted from dc model [9].

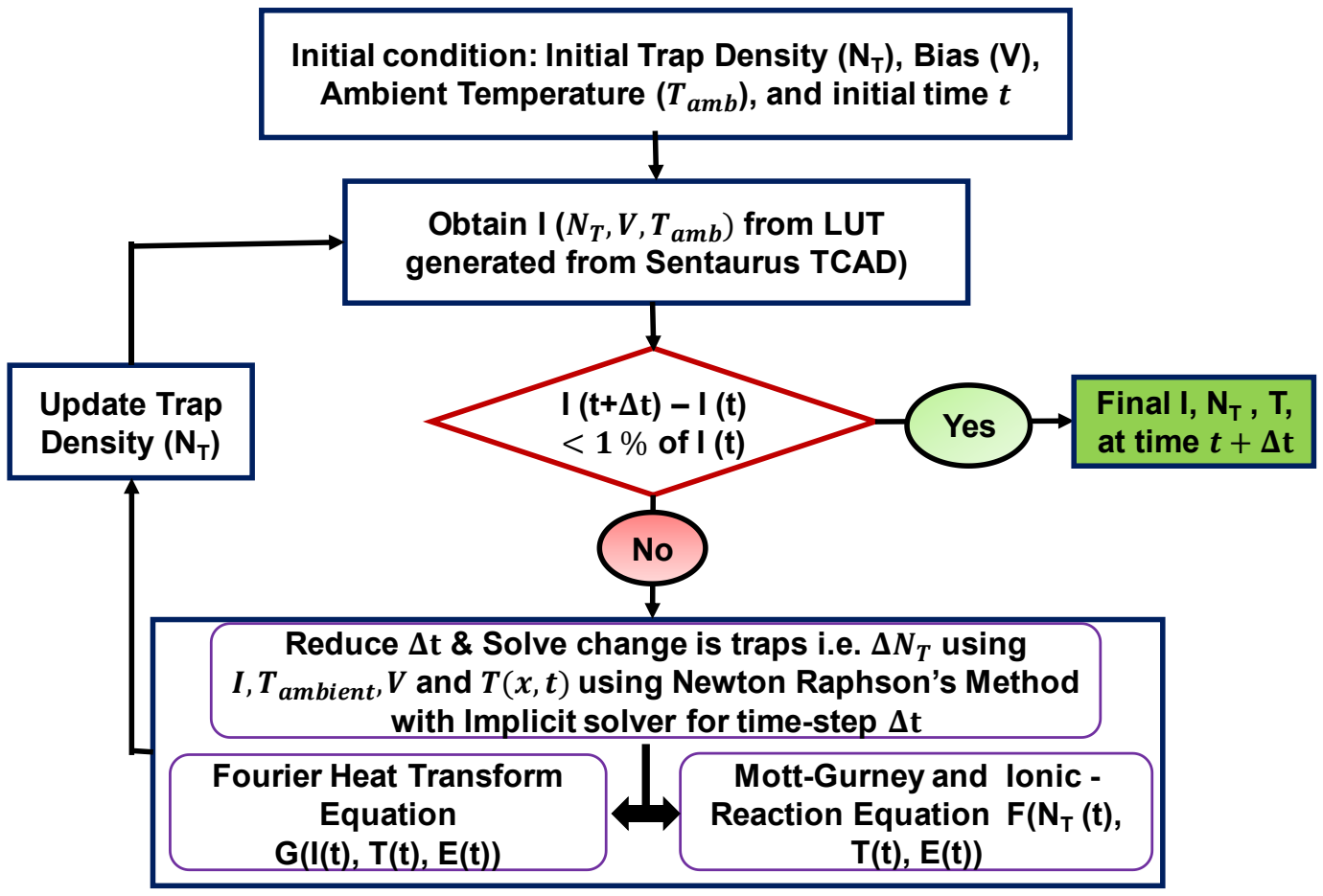

Fig. 3. Flowchart of Simulator for calculating Nt and I after a small time step $\Delta t$. This cycle is repeated to construct the entire transient of $10^{-8} - 1$s

### B. Solver

The numerical solver (Fig. 3) calculates the transient current, $I(t)$, by simultaneously solves a system of 3 equations in one dimension (1D) for simplicity, namely (i) Current transport ($I_q(t)$) (ii) Heat Flow to obtain temperature i.e. T(t) (iii) Ionic transport (reaction and drift) to obtain $N_T(x,t)$.

To decouple the hole current calculation from ionic transport calculation, we use two simplifying assumptions.

i) Ionic transport may lead to non-uniform trap density. However, we approximate the non-uniform trap density from ionic transport equation by an equivalent uniform trap density i.e. $N_T(x,t) = N_T(t)$ for current calculation. This assumption is reasonable because, firstly, we have earlier shown by simulations that asymmetry in trap profile $N_T(x)$ only weakly affects the largely symmetric IV characteristics [9]. Secondly, the experimental I(V) characteristics are essentially symmetric for positive vs. negative bias before the onset of ionic transport i.e. SET/RESET [9]. Hence, this provides a reasonably accurate estimation of current response to $N_T$ modulations without increasing the complexity significantly.

ii) The hole current response time is very fast compared to ionic current. Hence, hole current can respond instantaneously to change in trap density i.e. $N_T(t)$.

These two assumptions ensure that the current transport calculation simplifies to an equivalent function of uniform $N_T(t)$, ambient temperature, $T_{amb}$ and bias V(t) i.e. $I(N_T, T_{amb}, V)$ without detailed $N_T(x)$ profile considerations. Thus, we generate a look-up table (LUT) of $I(N_T, T_{amb}, V)$ based on quasi-static simulations.

As shown in Fig.3, first, we use initial $T_{amb}$, V, and $N_T(t)$ at a given time $t$. Second, we then use LUT to compute $I(N_T, T_{amb}, V)$ as shown in Fig. 4. Third, we compute the ionic and heat transport equations using an implicit 1D coupled differential equation solver based on Newton Raphson's method, for an incremental time step $\Delta t$ to obtain updated $N_T(t + \Delta t)$. Fourth, we then use LUT to compute $I(N_T(t + \Delta t), T_{amb}, V)$. Fifth, to ensure linearity, $I(t + \Delta t) - I(t)$ is more than 1% of $I(t)$ then time step, $\Delta t$, is reduced. Else $I(t + \Delta t)$ is final. This processed is repeated to to generate the entire transient current.

### C. *Current Transport by TCAD*

In current transport, we use electron-hole drift-diffusion DD simulations to compute PCMO current at a given uniform trap density ($N_T$) and ambient device temperature ($T_{amb}$). We had earlier demonstrated excellent matching of temperature dependent current transport in PCMO based RRAM using trap SCLC model with self-heating [10] implemented in Sentaurus™ [22]. Essentially the model solves the Poisson, carrier continuity, carrier statistics, and heat transport equations self-consistently. Further, the dc model was extended to perform transient simulations where transient current for fast SET and RESET was modeled to show excellent match with experimental current transients before the on-set of ionic transport [11].

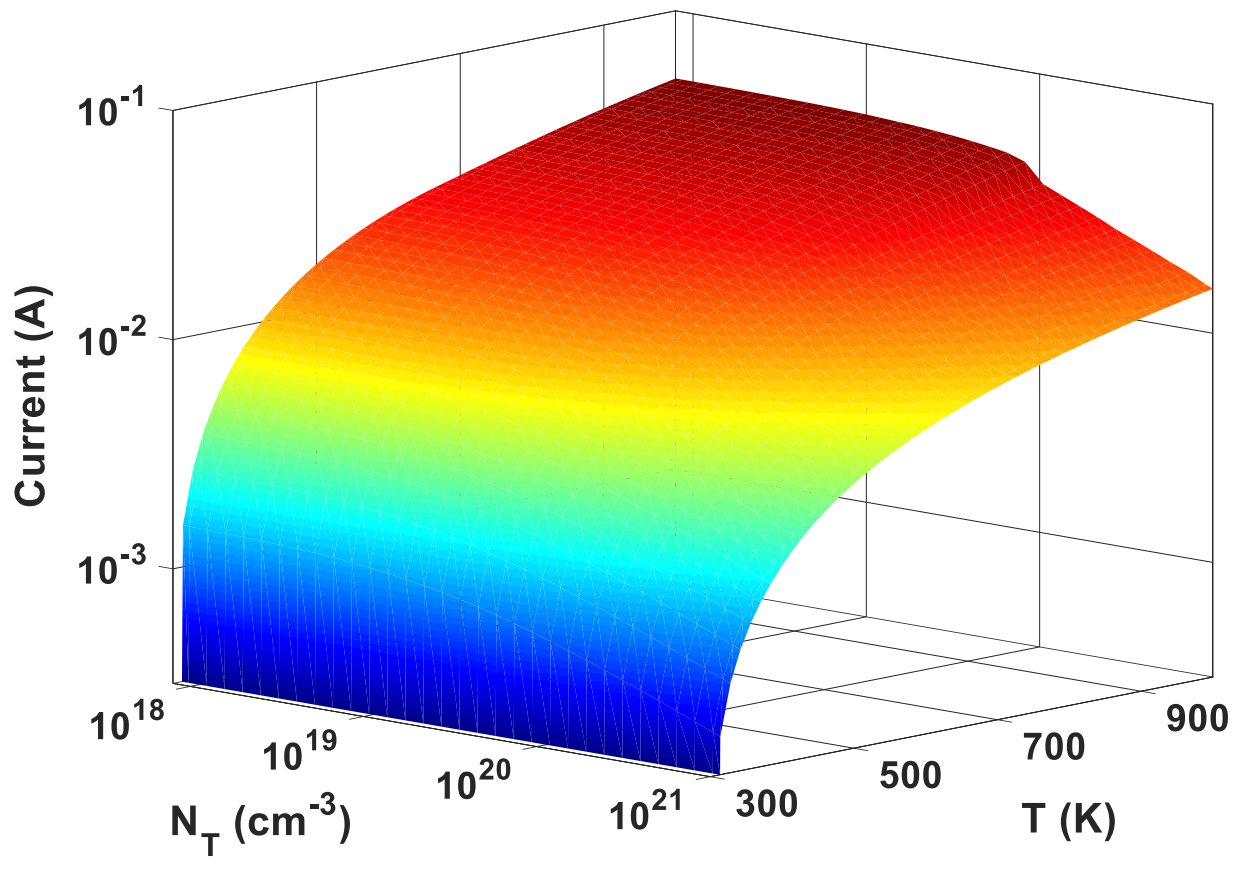

Fig. 4. Look- Up Table (LUT) for $I(N_T,\ T_{Peak})$ , $T_{amb} = 300K$

Thus, we have previously demonstrated a robust Sentaurus TCAD based solver for trap SCLC with self-heating. Based on this platform, we perform the quasi-static simulations in Sentaurus where we calculate $\mathrm{I}(\mathrm{N_T}, \mathrm{T_{amb}}, V)$. The internal device temperature is simultaneously calculated. Thus, LUT is integrated into the simulator in MATLAB$^{\mathrm{TM}}$, where we simultaneously solve the thermal and ionic motion equations as described in the following sections.

### D. *Thermal Model*

The 1-D Fourier heat transfer equation is solved for the device structure shown as follows.

$$-k\frac{d^2T}{dx^2} + c_v\frac{dT}{dt} = \frac{\mathrm{I}.V}{\mathrm{volume}} \qquad (12)$$

The thermal resistances of the top and bottom electrodes are calculated using FEM modeling as presented in detail in [10], [11]. The thermal conductivity of PCMO at 300K is 0.5 Wm$^{-1}$ K$^{-1}$ [23]. We included temperature dependence of heat capacity and thermal conductivity in simulations [24], [25].

The parameters used in the equations 1-12 are shown in Table 1.

TABLE I
PARAMETERS USED IN THIS PAPER

| Symbol | Quantity | Value | Reference |
|---|---|---|---|
| $E_M$ | Hopping barrier energy | 1.0 eV | [16] |
| a | Hopping distance | 1 nm | [16] |
| f | Escape frequency | $10^{13}$ Hz | [16] |
| $c_v$ | PCMO heat capacity | $2.76 \times 10^6$ J m$^{-3}$ K$^{-1}$ | [26] |
| $k$ | Thermal conductivity | 0.5 W m$^{-1}$ K$^{-1}$ | [23] |

## IV. RESULTS AND DISCUSSIONS

### A. *Effect of n on RESET transient*

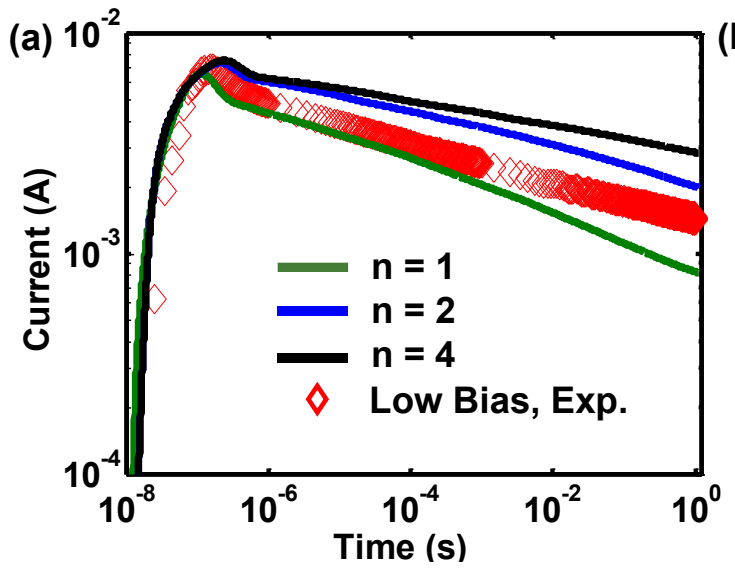

(b) TABLE II
SLOPE VALUES FOR n = 1,2,4

| n | Slope Without SH | Slope With SH |
|---|---|---|
| 1 | $-\frac{1}{2}$ | $-\frac{1}{7.7}$ |
| 2 | $-\frac{1}{3}$ | $-\frac{1}{10}$ |
| 4 | $-\frac{1}{5}$ | $-\frac{1}{15}$ |

Fig. 5. (a) Reset current vs time shows slow switching with different slop for n=1,2,4 (b) Different slope values during slow switching for n=1,2,4 with and without self-heating (SH)

The RESET transient for $\mathrm{n} = 1,2,4$ is shown in Fig. 5. Essentially, without self-heating, we obtain a $\mathrm{m} = -1/(\mathrm{n}+1)$ time exponent (not shown). However, with self-heating, the time exponent is reduced as shown in Fig. 5(b).

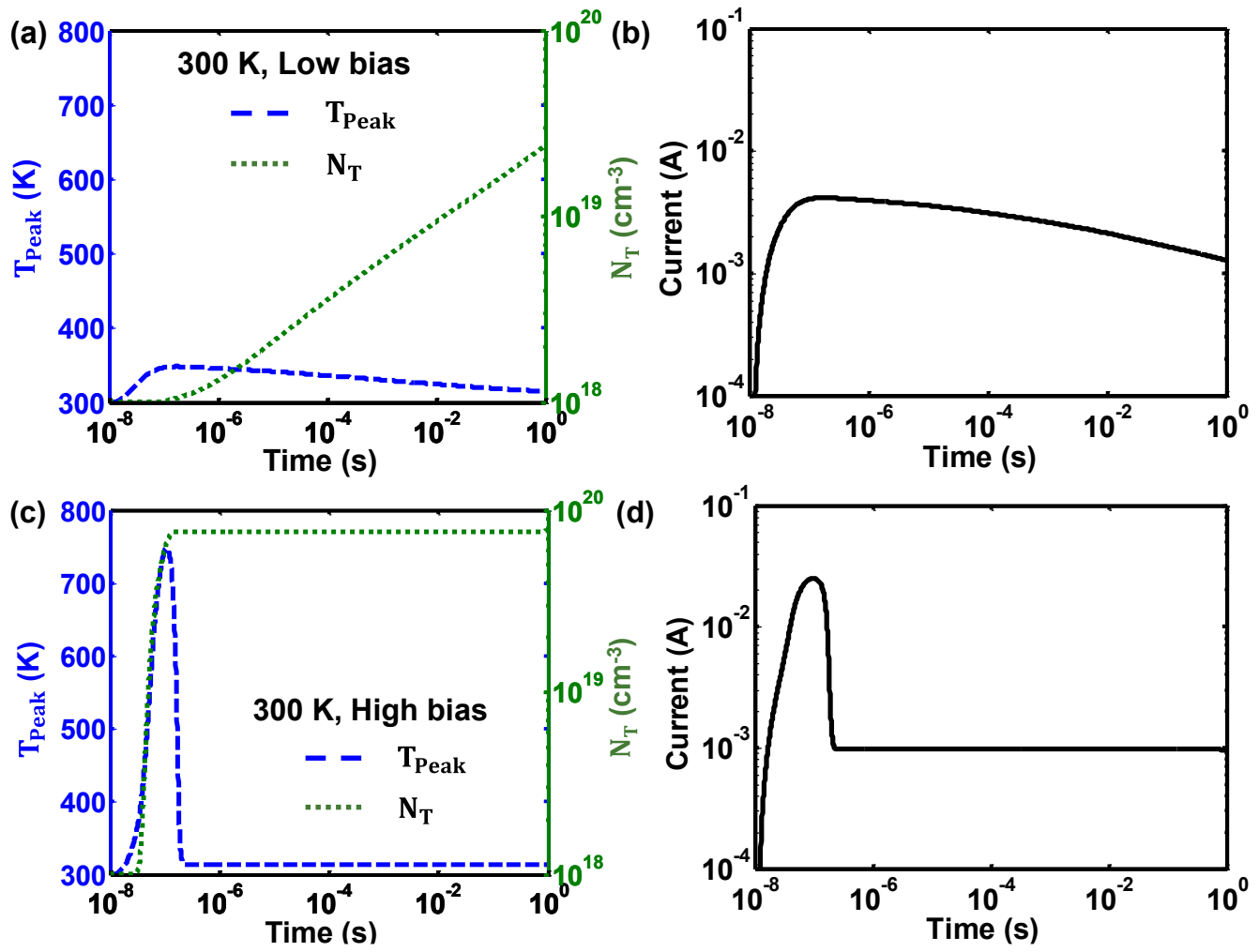

Fig. 6. Simulated (a) Temperature vs Time and Trap density vs time for low bias (b) Current vs time shows slow switching for applied low bias (c) Temperature vs Time and Trap density vs time for high bias (d) Current vs time shows fast switching for applied high bias.

The effect of bias on current, temperature and trap transients are shown in Fig. 6. At high bias, the three regimes are observed i.e. (i) current increases quickly and (ii) then decreases quickly (iii) to a settle to a saturation state. $T(t)$ increases quickly and then decreases to a saturation. $N_T(t)$ increases quickly and then saturate at low temperature (~320 K). At low bias, two regimes are observed (i) current increases and then (ii) follows the universal time exponent of $m = -1/10$. Here $T(t)$ also increases quickly but to a lesser extent than for high bias case. Then $N_T(t)$ increases slowly while $T(t)$ reduces slowly without saturation.

### *B. RESET Simulation*

Fig. 7 (a) compares simulated vs. experimental $I(t)$ for a range of bias. Our model is able to reproduce the experimental behavior quite comprehensively. Small quantitative differences can be attributed to the various simplifying assumptions detailed in Section II.

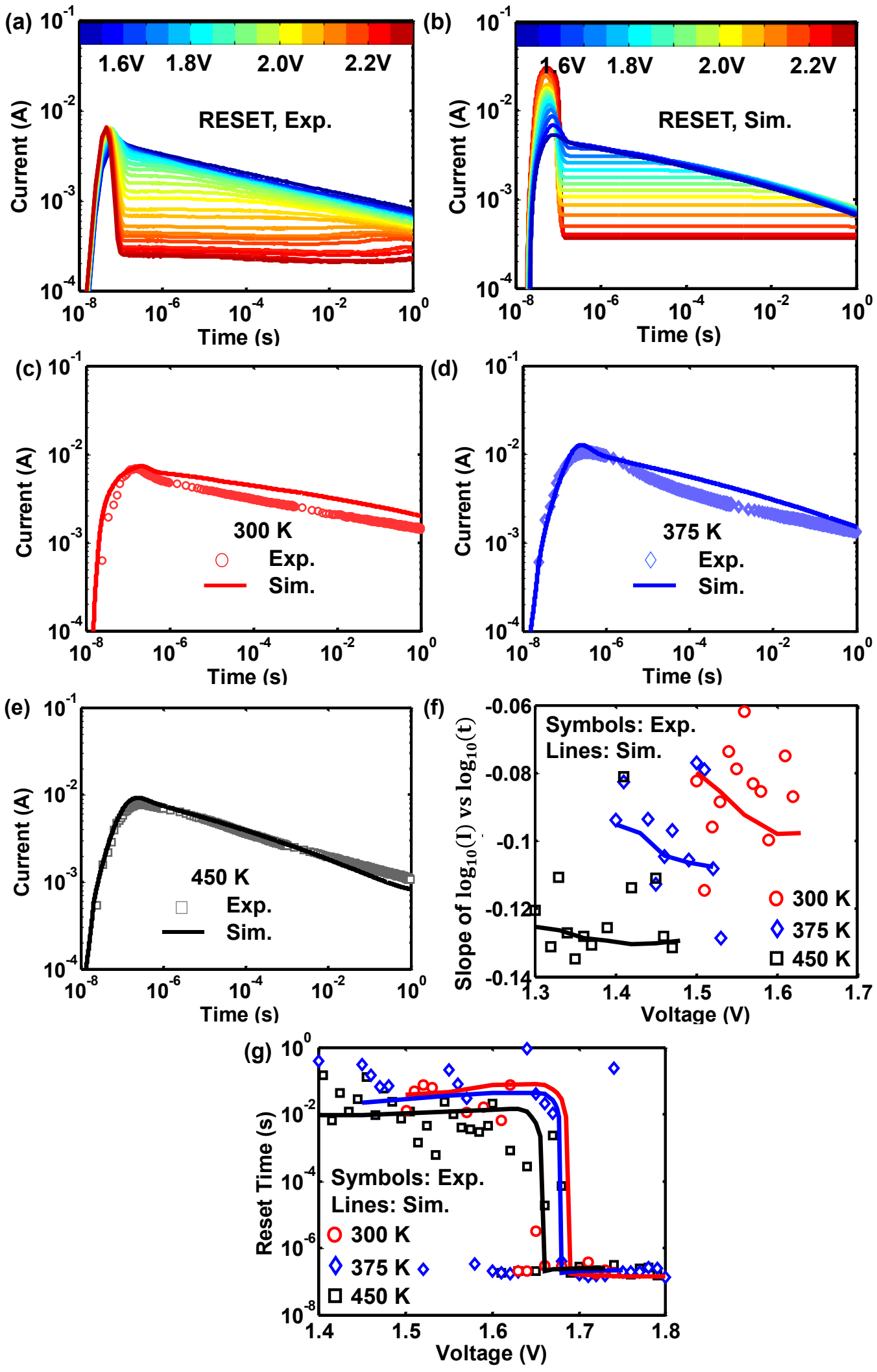

Fig. 7. RESET, (a) Experiment (b) Simulation shows qualitative matching of Current vs Time during RESET for range of step voltage input. (c), (d), (e) Matching of slow switching of current for different $T_{amb}$ (300-450K). (f) Matching of slow switching slope for a bias range $T_{amb}$ (i.e. 300-450K). (g) Reset time vs Voltage for three different $T_{amb}$ (i.e. 300-450K).

Next, we simulate the effect of ambient temperature ($T_{amb}$). Our transient simulations are in excellent agreement for a range of $T_{amb}$ (i.e. 300-450K). We have experimentally observed that the universal time exponent is weakly dependent upon $T_{amb}$, which is successfully captured in the simulations as shown in Fig. 7 (c-e). Fig. 7 (f) shows that the temperature and bias dependence of universal time exponent is captured well by simulations. Finally, the experimentally voltage-time dilemma is plotted in Fig. 7 (g) where the time to a fixed current level (i.e. 1mA) is extracted as switching timescale and plotted vs. applied bias. Simulations are presented for comparison. First, we observe that at high-bias, there is a saturation in switching speed to indicate that higher bias does not produce faster switching beyond 100ns timescale. Further, at low bias, the switching timescale is also limited 10ms. $T_{amb}$ does not strongly affect these levels. Simulations show excellent overall agreement with experiments.

### *C. SET Simulations*

We present SET transient for current, traps density and temperature at low and high bias and compare with simulation in Fig. 8. For low bias, current increases then saturate then increases to compliance. $T(t)$ shows a similar behavior while $N_T(t)$ initially remains high and then decreases abruptly as $T(t)$ increases to compliance. At higher bias, the stage S2 is shorter or non-existent.

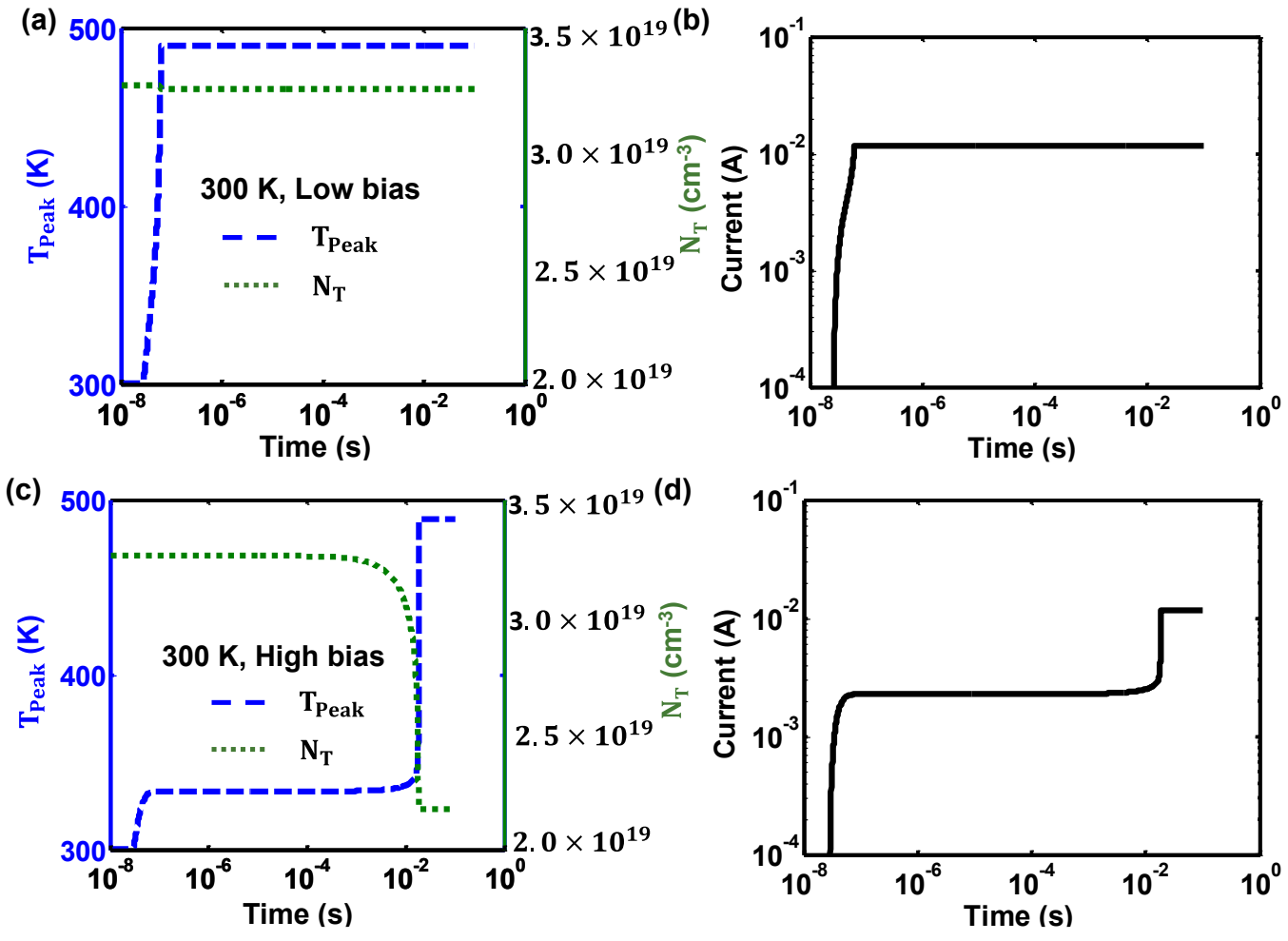

Fig. 8. Simulated (a) Temperature vs Time and Trap density vs time for low bias (b) Current vs time shows slow switching for applied low bias (c) Temperature vs Time and Trap density vs time for high bias (d) Current vs time shows fast switching for applied high bias.

Next, we show that the experimental SET transient (Fig. 9 (a)) is captured well by simulations at 300K seen in Fig. 9 (b). A time intercept at constant current of $10^{-2}$A is used as the estimate of SET time. Thus SET time vs. applied bias for $T = 300K, 375K, 450K$ is plotted in Fig. 9 (c). At higher bias, SET time shows SET time saturation and is essentially is limited to 100ns. At lower bias, there is a strong (~exponential) increase in SET time, which is essentially the voltage-time dilemma. This is in excellent agreement with simulations. Further, the $T_{amb}$ dependence of the SET behavior shows that the voltage-time dilemma for switching at low bias is aided by ambient temperature. However, at high bias, the timescale of SET is limited to 100ns, which is temperature insensitive. These experimental results are also in excellent agreement with simulations.

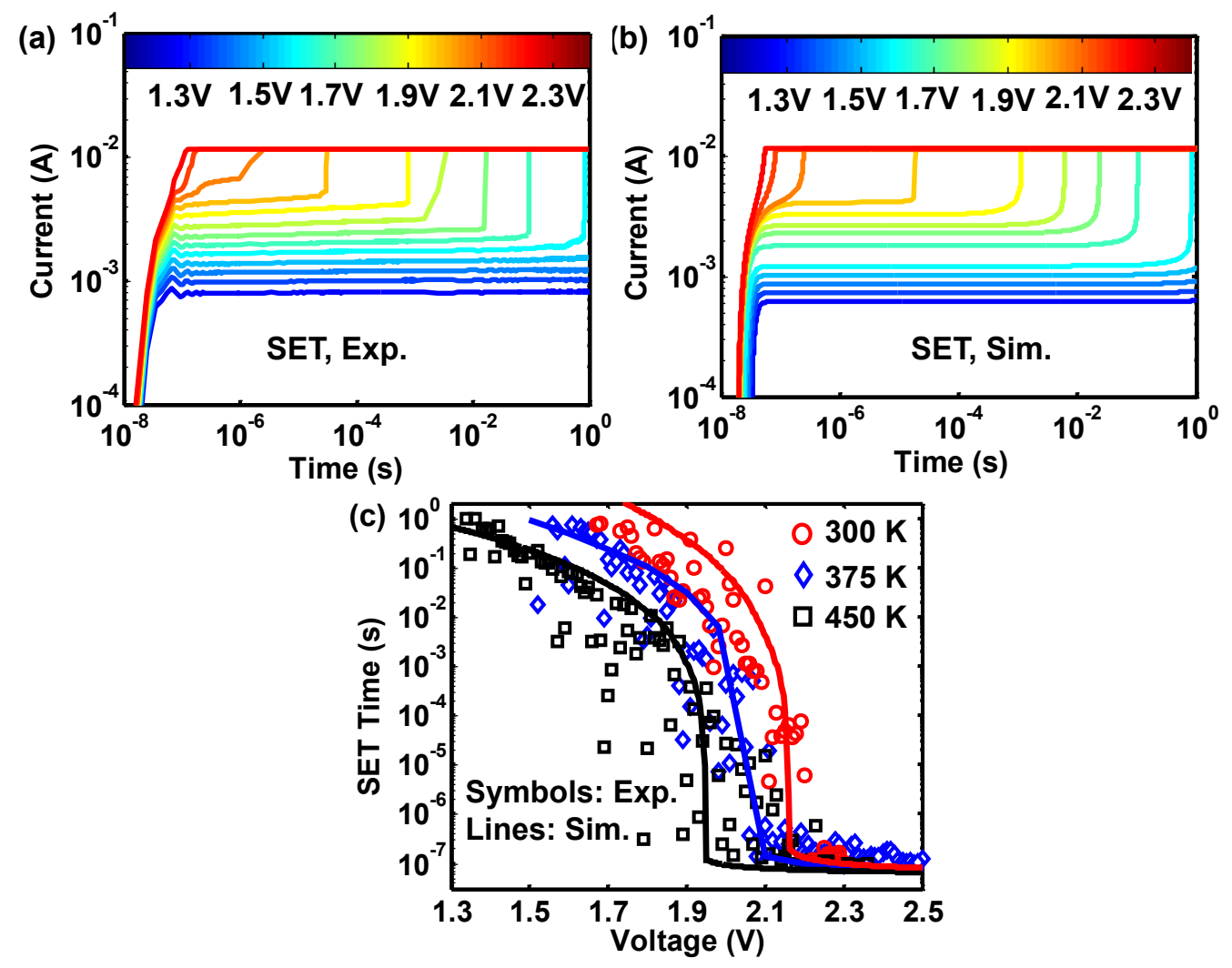

Fig. 9. SET, (a) Experiment (b) Simulation shows qualitatively matching of Current vs time during SET for different step voltage input. (c) SET time vs Voltage for three different $T_{amb}$ (i.e. 300-450K).

## V. CONCLUSION

In this paper, a Reaction Drift Model is proposed to include ion dynamics to a drift-diffusion with self-heating based model for hole transport. We demonstrate that the model can reproduce experimentally observed SET and RESET transient across a range of timescale $(10^{-8} - 1s)$, SET/RESET bias and ambient temperatures $(300 - 450K)$. Remarkably, a universal RESET behavior is of time-exponent of $m \approx -1/10$ is replicated. The simulations are able to capture the difference between SET-RESET timescale vs SET-RESET voltage to explain the different voltage-time dilemma observed in SET vis a vis RESET. The ambient temperature shows a stronger effect for SET compared to RESET – which is captured well in simulations. Further, the timescale for fast switching is limited to ~100ns for SET and RESET – which is independent of ambient temperature is also captured. Thus, we present a simple 1D model of SET /RESET in PCMO based RRAM that can comprehensively reproduce timescale, bias and temperature effects. Such a model will enable a detailed understanding and design of PCMO based RRAMs.